%% file: acceleration_paper-initial.tex
\documentclass{pasa}%
\pdfoutput=1
\usepackage{graphicx}

\title[Stellar Accelerations and the Galactic Gravitational Field]{Stellar Accelerations and  the \\ Galactic Gravitational Field}

\author[Silverwood and Easther]{H. Silverwood$^1$, R. Easther$^2$
\affil{$^1$Institut de Ci\`encies del Cosmos (ICCUB), Universitat de Barcelona (IEEC-UB), Mart\`i Franqu\`es 1, E08028 Barcelona, Spain}%
\affil{$^2$Department of Physics, University of Auckland, Private Bag 92019, Auckland, New Zealand}
}%

\jid{PASA}
\doi{10.1017/pas.\the\year.xxx}
\jyear{\the\year}

\usepackage{aas_macros}
\usepackage{hyperref} 
\hypersetup{colorlinks,citecolor=blue,linkcolor=blue,urlcolor=blue}

\usepackage[usenames, dvipsnames]{color}
\usepackage[caption=false]{subfig}
\usepackage{amsmath}    
\usepackage{nicefrac}

\newcommand{\Msun}{{\rm M}_\odot}

\newcommand{\cms}{{\rm cm\,s}^{-1}}
\newcommand{\ms}{{\rm m\,s}^{-1}}
\newcommand{\mss}{{\rm m\,s}^{-2}}
\newcommand{\kms}{{\rm km\,s}^{-1}}

\newcommand{\muas}{\mu {\rm as}}
\newcommand{\muasyr}{\mu {\rm as\,yr}^{-1}}
\newcommand{\nasyr}{{\rm nas\,yr}^{-1}}

\def \eg{{\textit{e.g.~}}}
\def \ie{{i.e.~}}
\newcommand{\rmd}{{\rm d}}
\input{jdefs.tex}
\hypersetup{draft}

\begin{document}

\begin{frontmatter}
\maketitle

\begin{abstract}
Typical stars in the Milky Way galaxy have velocities of hundreds of kilometres per second and experience gravitational accelerations of $\sim10^{-10}~\mss$, resulting in velocity changes of a few centimetres per second over a decade.  Measurements of these  accelerations  would permit direct tests of the applicability of Newtonian dynamics on kiloparsec length scales  and could reveal significant small scale inhomogeneities within the galaxy, as well increasing the sensitivity of measurements of the overall mass distribution of the galaxy.   Noting that a reasonable extrapolation of progress in exoplanet hunting spectrographs suggests that centimetre per second level  precision will be attainable in the coming decade(s), we explore the possibilities  such measurements would create.  We  consider possible confounding effects, including apparent accelerations induced by stellar motion and reflex velocities from planetary systems, along with possible strategies for their mitigation. If these issues can be satisfactorily addressed it will be possible to use high precision measurements of changing stellar velocities  to perform a ``blind search'' for dark matter,  make direct tests of theories of non-Newtonian gravitational dynamics, detect local inhomogeneities in the dark matter density, and greatly improve measurements of the overall properties of the galaxy.
\end{abstract}

\end{frontmatter}

\section{INTRODUCTION } \label{sec:intro}
The dynamics of galaxies constitute one of the most profound challenges faced by astrophysics.  This is a long-standing problem: suggestions that the gravitational dynamics of the Milky Way galaxy could be dominated by nonluminous matter predate  confirmation that the Milky Way is itself one of many galaxies in the universe \citep{kelvin_1904}. Conversely, appeals to ``modified gravity'' at galactic length scales undermine the bedrock assumption that the laws of physics revealed by terrestrial experimentation apply equally  on astrophysical scales -- a paradigm with its origins in Newton's explanation of the Keplerian dynamics of the solar system. It is thus hard to overstate the potential impact of this topic.

On scales of light years and above, however, we  can  never directly recover the local gravitational force;   observables are positions, velocities (from  redshifts or stellar proper motions) and integrated quantities such as photon trajectories, the outcome of structure formation, or microwave background temperature anisotropies.  Forces  are proportional to accelerations and measuring these directly requires the ability to discern  changes in the velocities of objects over time. Given that  stellar velocities are hundreds of kilometres a second in the galactic rest-frame  at radii of $\sim$10 kpc, we immediately estimate that typical stellar accelerations are $\sim 10^{-10}~\mss$, corresponding to a change in velocity of a few centimetres per second per decade. 

Spectrographs now under construction aim to  measure stellar radial velocities at precisions of  ${\cal O}(10~\cms)$ \citep{2017AAS...22912604F}. Exoplanet searches provide the primary motivation for these ultra-high resolution instruments; an Earth-like planet orbiting a Sun-like star induces a velocity of roughly 10 $\cms$ in the star. Given that Doppler shifts are proportional to $v/c$, this is a parts-in-a-billion measurement of the stellar spectrum -- a stunning technical achievement. However, it is also the latest step in a decades-long sequence of instruments. The challenges of continuing this rate of progress should not  be underestimated but it is plausible that  ``next-to-next'' generation  instruments will be capable of measuring stellar accelerations due to the galactic gravitational field over a baseline of several years.

The present paper explores this possibility, and examines possible issues with its  implementation.  Such measurements would constitute a galactic analogue to the Sandage-Loeb effect \citep{1962ApJ...136..319S,1998ApJ...499L.111L}. Just as resolving the changing expansion rate of the universe would provide novel constraints on the dynamics and composition of the universe as a whole, direct measurements of stellar accelerations  open new lines of enquiry into the structure of the galaxy and the fundamental properties of gravity. In particular, these would include direct tests of the validity of Newtonian dynamics on kiloparsec length scales.

Clearly, any feasible timeline for the full realisation of this strategy is measured in decades rather than years. However, direct measurements of the acceleration field can map  the galactic mass distribution and probe the  substructure of the Milky Way in  novel ways. Moreover, this strategy  provides an  new approach to disentangling modified gravity from dark matter, a problem over a century old.  Given the vast range of possible dark matter scenarios -- each with its own parameter space -- it is entirely conceivable (and perhaps even likely from some standpoints) that dark matter is effectively undetectable by {\em any\/} plausible instrumentation or experiment, even if the concordance cosmology is fundamentally correct.\footnote{Recall that the fundamental dynamic range of potential dark matter ingredients runs from the $10^{-22}$ eV of ultralight dark matter through to multisolar mass black holes; specific models can be tightly constrained but our integrated experimental effort only explores a tiny range of the total parameter space.}  

There is no fundamental connection between the radial acceleration induced by terrestrial planets orbiting sunlike stars and galactic dynamics, so it perhaps unsurprising that little attention has been paid to the fact that  exoplanet searches based on Doppler spectroscopy are approaching the regime where these measurements become possible. Moreover, as modified gravity theories typically contain a fundamental acceleration parameter $a_0$, which is necessarily of the same order as the accelerations experienced by stars moving in a galactic potential \citep{1983ApJ...270..365M},  this target is one that has been clearly visible in the literature. That said, \citet{2008MNRAS.391.1308Q} proposes mapping the Milky Way potential via the peculiar accelerations of  globular clusters, but its authors are less than optimistic about the immediate applicability of the technique. We also draw attention to \citet{HarvardMichiganPaper}, which provides a complementary analysis of a similar proposal to the one presented here, and which was completed contemporaneously with this work. 

Computing both line-of-sight and transverse accelerations shows that direct measures of transverse accelerations would require a substantially greater advances in astrometric precision than the corresponding improvement needed in spectroscopic measurements of radial velocities. However, these two effects are coupled via perspective acceleration, as the changing viewing angle associated with stellar proper motion  changes the component of the velocity vector parallel to the line of sight. The magnitude of this effect decreases with distance but can easily swamp the gravitational acceleration. Fortunately,  this effect can be estimated and accounted for with suitable precision by next generation astrometric measurements of proper motion and parallax.

Having argued that it will apparently be feasible to  directly measure stellar accelerations, we consider two specific physical scenarios we could investigate with this data. Significant, localised dark matter overdensities or other substructure can be revealed by direct measurements of local accelerations. Furthermore, assuming that gravity (outside of lensing effects) is well-described by Newtonian dynamics, stellar acceleration information could provide a direct, tomographic measurement of the mass-distribution in the Milky Way galaxy.  Secondly, this work facilitates detailed comparisons of MOND and Newtonian dynamics. Their predictions of  standard and modified gravity overlap in the plane of the galaxy (by design, since modified gravity was built to account for these observations) but  predictions for motion in the vertical direction generically differ. 

In what follows we explore strategies to obtain and make use of acceleration information. In  Section \ref{sec:Milky Way_acceleration_field} we map out the acceleration field of the Milky Way with a conventional disc+halo model and compute expected radial and transverse accelerations.  In  Section \ref{sec:spectroscopic_performance} we consider the practicalities of implementing this proposal in terms of stellar magnitudes, spectral types, and the likely roadmap for future instrumentation.  In Section~\ref{sec:perspective_acceleration} we show that perspective accelerations from changing lines of sight resulting from proper motion are similar to or larger than that derived from the Milky Way potential, but that high-precision astrometric measurements of parallax and proper motion can correct for this effect. In Section \ref{sec:exoplanetary_background} we explore strategies for extracting the galactic signal from accelerations  induced by stellar planetary systems.   Finally, in Section~\ref{sec:disc_morphology} we discuss the ability of these measurements to detect large-scale inhomogeneities in the dark matter distribution and to probe large scale gravitational physics. We discuss our results in Section \ref{sec:conclusions}. 

\section{Milky Way Acceleration Field }\label{sec:Milky Way_acceleration_field}

The Poisson equation links the density $\rho$, the Newtonian potential $\Phi$ and local gravitational acceleration $\vec{g}$:
\begin{equation}
4 \pi G \rho = \nabla^2 \Phi = - \nabla \cdot \vec{g}, 
\end{equation}
where $G$ is the gravitational constant.  We can investigate the mass distribution and potential of the Milky Way if we can measure the accelerations of a set of tracer stars at positions $\vec{x}_i$ given by $\vec{g}(\vec{x}_i)$. Currently, only  positions and velocities $\vec{v}_i$ are measured.  The direct use of accelerations -- or even a single component of the full 3D acceleration -- would remove the need for many of the assumptions and approximations used when recovering the potential with methods such as the collisionless Boltzmann equation in combination with distribution function modelling \citep[\eg][]{2015MNRAS.454.3653B} or Jeans equation modelling (\citealp{1922MNRAS..82..122J, 2016MNRAS.459.4191S,  2018MNRAS.478.1677S}, see \eg  \citealp{2014JPhG...41f3101R} for an overview), and reveal a more direct and unbiased view of the Milky Way's potential and mass distribution.

\begin{figure}
\begin{center}
\includegraphics[width=\columnwidth]{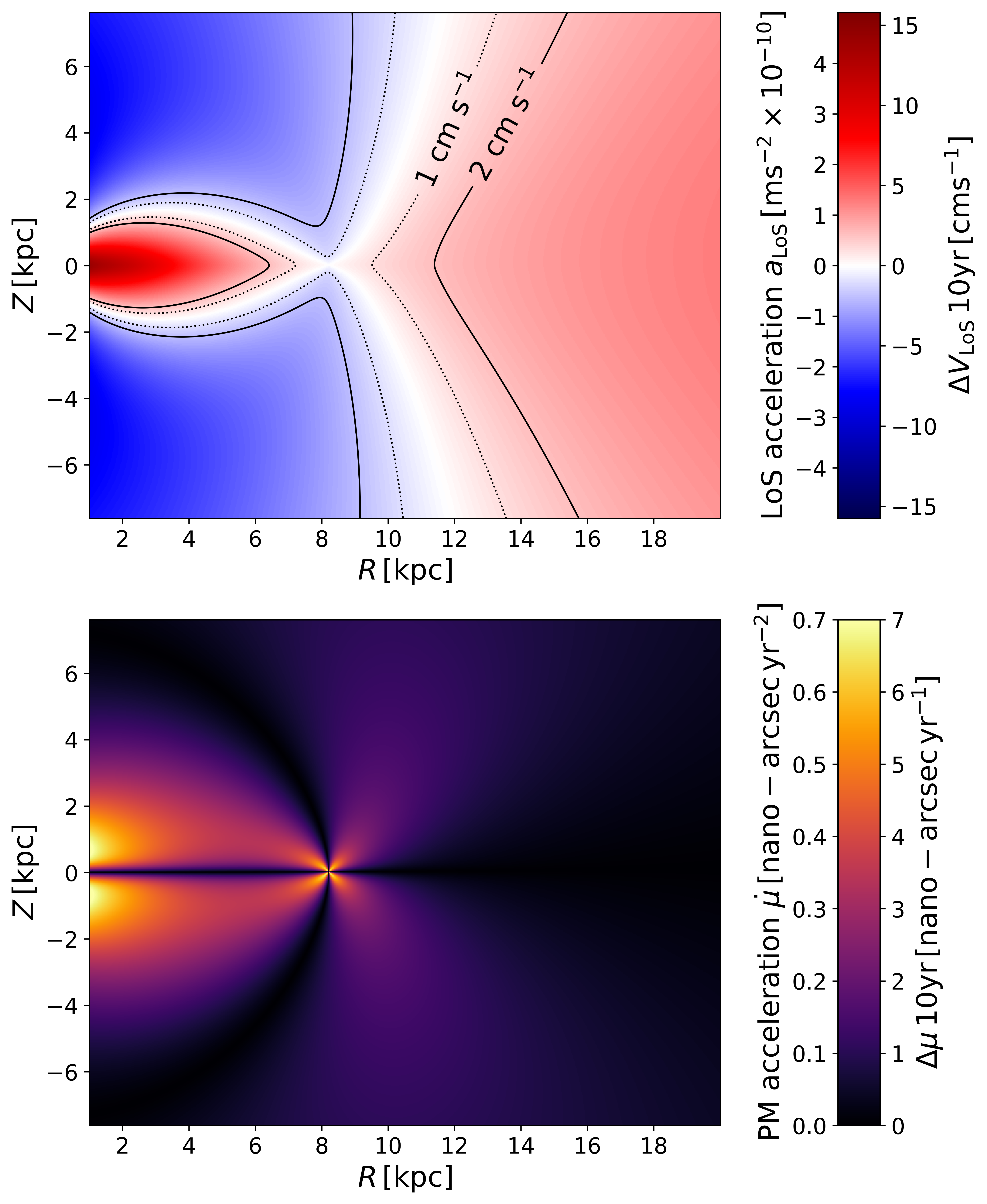}
\caption{Line of sight (top) and proper motion (bottom) heliocentric accelerations via {\tt Milky WayPotential2014}   \citet{2015ApJS..216...29B}, across the $Rz$ plane of the Milky Way. Accelerations are indicated by the left-hand scales on the colourbars; total change over a decade is shown on the right hand scale.   In the top plot  dotted and solid lines indicate the 10 year  $\Delta V_{\rm LoS} = 1\cms$ and $\Delta V_{\rm LoS} = 2\cms$ contours, respectively. The latter $\Delta V_{\rm LoS}$ is the sensitivity goal of the planned CODEX spectrograph \citep[][see Section \ref{sec:spectroscopic_performance}]{2010SPIE.7735E..2FP, 2010Msngr.140...20P}.}
\label{fig:mw_accelerations}
\end{center}
\end{figure} 

To begin our discussion we map the acceleration field of the Milky Way, taking into account the acceleration of the Sun. We work with the {\tt MilkyWayPotential2014}   model  in the {\tt galpy} package \citep{2015ApJS..216...29B}. Radial and vertical accelerations at each point in the $Rz$ plane are converted into heliocentric values by subtracting the acceleration at the solar position ($R=8.2$kpc, $z=25$pc \citet{2016ARA&A..54..529B}). The galactocentric radial and vertical accelerations are projected along the line of sight (LoS) between the Sun and a given position to provide the line of sight acceleration $a_{\rm LoS}$. The vector rejection (the acceleration orthogonal to the line of sight) then gives the perpendicular acceleration $a_{\rm perp}$, which is converted to the proper motion acceleration via $\dot{\mu} = \tan^{-1} \left(a_{\rm perp} /d \right)$ where $d$ is the distance to the Sun.  The line of sight and proper motion accelerations are shown in Figure \ref{fig:mw_accelerations}. For convenience we also express the accelerations as a velocity change over a fiducial interval of decade.

For typical stellar separations, the impact of neighbouring stars is negligible compared to that of the overall galactic gravitational field. The Newtonian gravitational  acceleration $a$ on a star from another star of mass $m$ at distance $r$ is $a = Gm/r^2$. For a $1~\Msun$ star at 4 lightyears (or 1.3 pc, roughly the distance to Alpha Centauri) the acceleration is $\sim 10^{-13}~\mss$, far smaller than the signal we would be seeking to detect. 

Within a kiloparsec of the Sun we find regions where the 10 year change in velocity, $\Delta V_{\rm LoS} \sim 1~\cms$. Within this range, stars in the vertical direction have higher accelerations, reflecting the steeper potential gradient moving away from the baryonic disc. At greater distances toward the Galactic Centre,  there are regions where $\Delta V_{\rm LoS} \sim 10~\cms$ at galactocentric radii of $\lesssim 3$ kpc.

The changes in proper motions over a 10 year period are $\mathcal{O}(\nasyr)$\footnote{The abbreviation ``nas'' denotes  nano-arcseconds.}. To put this into context, the Gaia mission has at best a proper motion sensitivity on the order of $10~\muasyr$ \citep{2018A&A...616A...1G}, along with parallax precision of $\mathcal{O}(10~\muas)$. Looking ahead, Gaia-NIR, a proposed successor mission with a similar scanning strategy to Gaia but observing in the near-infrared would have comparable sensitivity, and when combined with data from Gaia could yield a factor 14 improvement in proper motion precision due to the $\sim 20$ year time baseline between the  missions \citep{2016arXiv160907325H,2018IAUS..330...67H}, and double the parallax precision \citep{2014arXiv1408.2190H}. The proposed Theia mission would have at best $\mathcal{O}$(100 nas) and $\mathcal{O}(100~\nasyr)$ parallax and proper motion precision respectively \citep{2017arXiv170701348T}. 

Sub-$\muas$ and nas precision astrometry was discussed as a goal for a large L-class mission \citep{CosmicVision2013, 2014EAS....67..307B}. Optical and infrared  astrometric precision scales as $\sigma \propto \lambda / (B \sqrt{N})$, where $\lambda$ is the wavelength of light observed, $B$ is the mirror aperture size of a single telescope system or the baseline distance of an interferometer system, and $N$ is the number of photons  collected \citep{2005ESASP.576...29L}. With current photon collection efficiency already near 100\% and most stars emitting in the optical or infrared bands, the only avenue for increasing precision is the mirror aperture or interferometer baseline $B$. \citet{CosmicVision2013} and \citet{2014EAS....67..307B} explored an interferometer mission with a collecting area of a few square metres and a baseline of 100-1000m, requiring  ``formation flying'' and considerable advances in   global astrometry. Thus $\nasyr$ proper motion precisions are imaginable,  but it will not be achieved in the near or even medium term. By contrast there is a concrete path toward $\mathcal{O}(\cms)$ precision radial velocities, so we focus on this approach in what follows.  

Separately from precision stellar astrometry and spectroscopy, pulsar timing is an increasingly important tool for high-precision, astrophysical measurements that extend beyond the properties of the pulsars themselves. Pulsars come in a wide variety of types, with millisecond pulsars providing the greatest stability, rivalling that of simple atomic clocks \citep{2009ApJ...693L.109K}. Individual binary systems allow direct tests of relativistic effects, including orbital decay via the emission of gravitational radiation \citep{1975ApJ...195L..51H,2016ApJ...829...55W} and pulsar timing arrays are expected to be sensitive to both point source and stochastic nanoHertz gravitational wave backgrounds in the coming decade \citep{2018arXiv181108826B}. 

To our knowledge, the use of a pulsar timing array as a probe of the overall galactic gravitational field has not been investigated in detail. However, timing data is certainly sensitive to local accelerations and has been used for exoplanet detection \citep[\eg][]{1992Natur.355..145W}. Likewise, pulsar timing measurements have already been used to measure the high accelerations found in extreme locations such as globular clusters or the vicinity of large black holes \citep[\eg][]{2014ApJ...795..116P, 2017MNRAS.468.2114P, 2017MNRAS.471.1258P, 2018MNRAS.473.4832G}.  Pulsars are subject to spin-down which will produce frequency changes several orders of magnitude greater than those produced by Milky Way acceleration field, so reliably removing this will be critical to the detection of any linear change in velocity \citep{1992RSPTA.341...39P,2017MNRAS.467.3493J}.  

As we discuss in Section~\ref{sec:perspective_acceleration}, the changing line of sight induced by relative motion between the pulsar and the Sun will typically need to be accounted for in order to reveal the galactic acceleration; pulsar timing data has been used to obtain astrometric data from the parallax acceleration, but requires a model of the gravitational field of the galaxy as input \citep{2016ApJ...818...92M}. This degeneracy can  be broken by direct VLBI measurements of pulsar parallax and proper motion and analogous measurements to those discussed here are  potential targets for future pulsar timing studies. 
 
\section{Spectroscopic Performance}\label{sec:spectroscopic_performance}
\begin{figure*}
\begin{center}
\includegraphics[width=\textwidth]{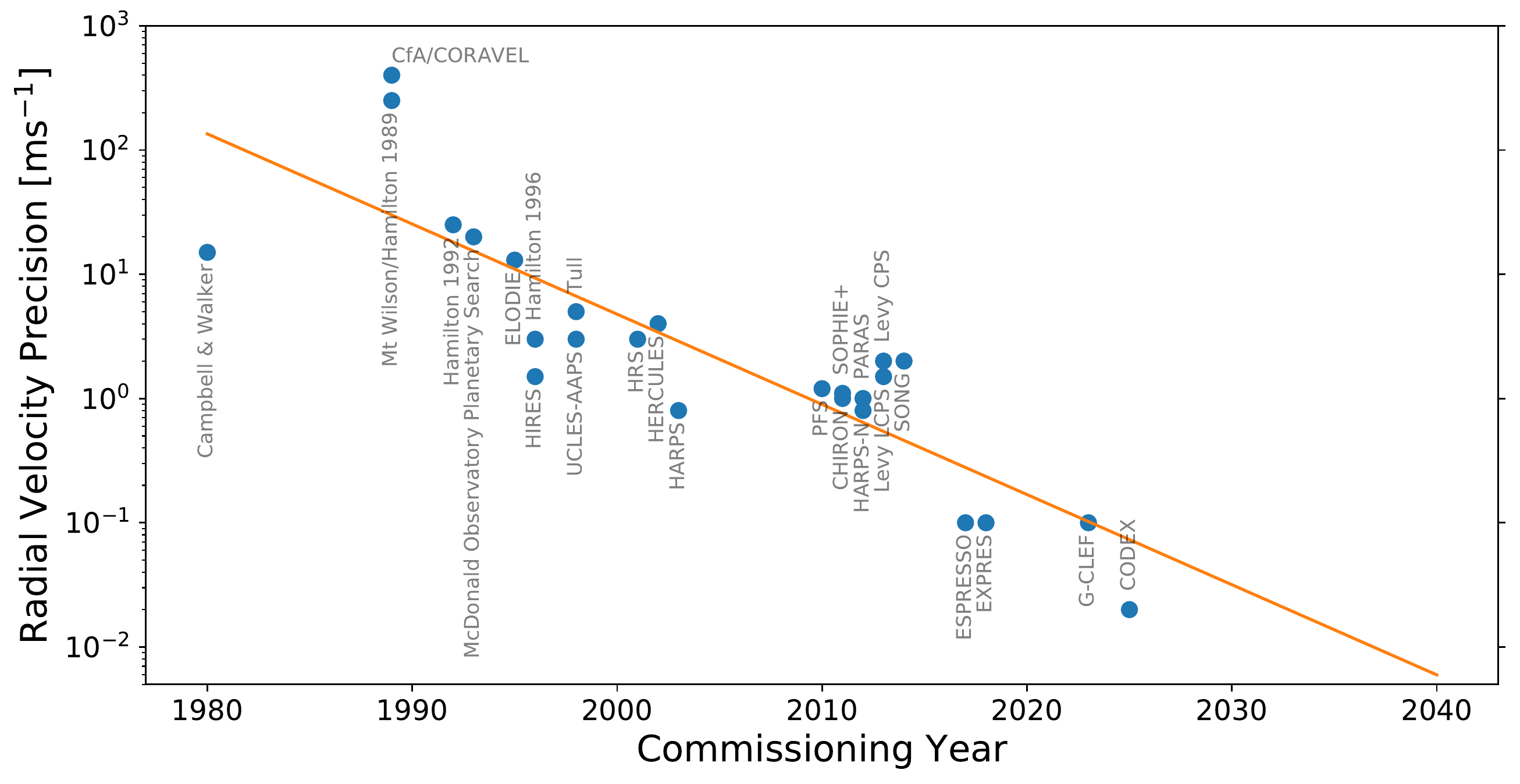}
\caption{Spectrographic precision since 1981, and an extrapolation to 2040. Drawn from data presented in \citet{2013pss3.book..489W} and \citet{2016PASP..128f6001F}. Full references for each spectrograph are listed in Appendix \ref{app:spectrograph_references}.}
\label{fig:spectrograph_precision}
\end{center}
\end{figure*}

Hunting exoplanets is a key motivation for improving spectrographic precision.  Modern precision spectrographs are vibrationally isolated,  temperature stabilised and in vacuum chambers.  Stable reference sources are needed to provide a time-independent calibration to track changes over long time intervals\footnote{With a single measurement Doppler shifted spectral lines can be compared to those measured in a laboratory to determine the \textit{absolute} radial velocity of a star. This is the general methodology of the radial velocity surveys used in galactic dynamics, which typically have precisions of $\mathcal{O}(100-1000~\ms)$ (\eg SDSS, \citealp{2011AJ....142...72E}; LAMOST, \citealp{2015MNRAS.448..822X}; RAVE, \citealp{2017AJ....153...75K}; and GALAH, \citealp{2018MNRAS.478.4513B}). Exoplanet hunting spectrographs extract higher precision \textit{relative} radial velocities in part by measuring instead the change in spectral lines over time, cancelling out some systematic uncertainties \citep{2017haex.bookE...4W}.}, for which the current state of the art is laser frequency combs linked to an atomic clock. See \eg \citet{2017haex.bookE...4W} and \citet{2016PASP..128f6001F} for reviews of spectrographic measurements.  

Figure \ref{fig:spectrograph_precision} shows the roughly log-linear progress in the precision of planet hunting spectrographs since 1980. Such Moore's Law-like progressions are not automatic, but reflect a deliberate, challenging, decades-long campaign to realise a series of significant technical improvements in spectrographic technology. However, it is not unreasonable to project that the $\mathcal{O}(\cms)$ velocity changes  resulting from the galactic gravitational field  will soon be within reach. 

Not all stars are suitable for high precision radial velocity measurements. If a star is too hot ($T_{\rm eff} > $10000 K) the absorption lines are smeared into a continuum, and if a star is too cold ($T_{\rm eff} < $3500 K) the lines can become too densely packed and overlapping to be distinguishable. Colder stars are also typically fainter, making it harder to achieve a specified signal to noise ratio \citep{2010exop.book...27L}.  

A stellar spectrograph measures the integrated output from the stellar surface, and stars with rapidly moving features in their photospheres exhibit \textit{jitter}, creating variation in their apparent Doppler velocities \citep[\eg][]{2001A&A...379..279Q}.  The impact of this jitter can be reduced by selecting stars for which these effects are typically small; ideal candidates are cool, old G- and K-dwarf stars \citep{2006A&A...454..943H,2013pss3.book..489W}. G- and K- dwarves show variations of $\mathcal{O}(100~\cms)$ \citep{2005PASP..117..657W}, while K giants have variations of $\mathcal{O}(10~\ms)$ \citep{2006A&A...454..943H}. Careful modelling can separate jitter from true stellar motion as the physical mechanisms that produce jitter are correlated with variations in the detailed line shapes and profiles \citep[see \eg][for a full overview]{2017haex.bookE...4W, 2017ApJ...846...59D}. 

The same photospheric features that cause jitter in radial velocities can also generate astrometric jitter, impacting precision parallax and proper motion measurements \citep[see \eg][and references therein]{2007A&A...476.1389E}. For instance a dark spot on the surface of a rotating star will shift the observed radial velocity up or down, depending on whether it is moving towards or away from us. Astrometric methods rely on measuring the location of the photocentre of a star, and the presence of a dark spot will shift this photocentre towards the opposite side of the star's visible disc. As dark spots form, rotate, and/or dissolve the photocentre will shift, perhaps mimicking accelerations acting on the star. Due to their common origin radial velocity and astrometric jitter are correlated \citep{2007A&A...476.1389E}, opening a further area of complementarity between high precision radial velocity and astrometric measurements that might be exploited in  future work.      
 
\begin{figure}
\includegraphics[width=\columnwidth]{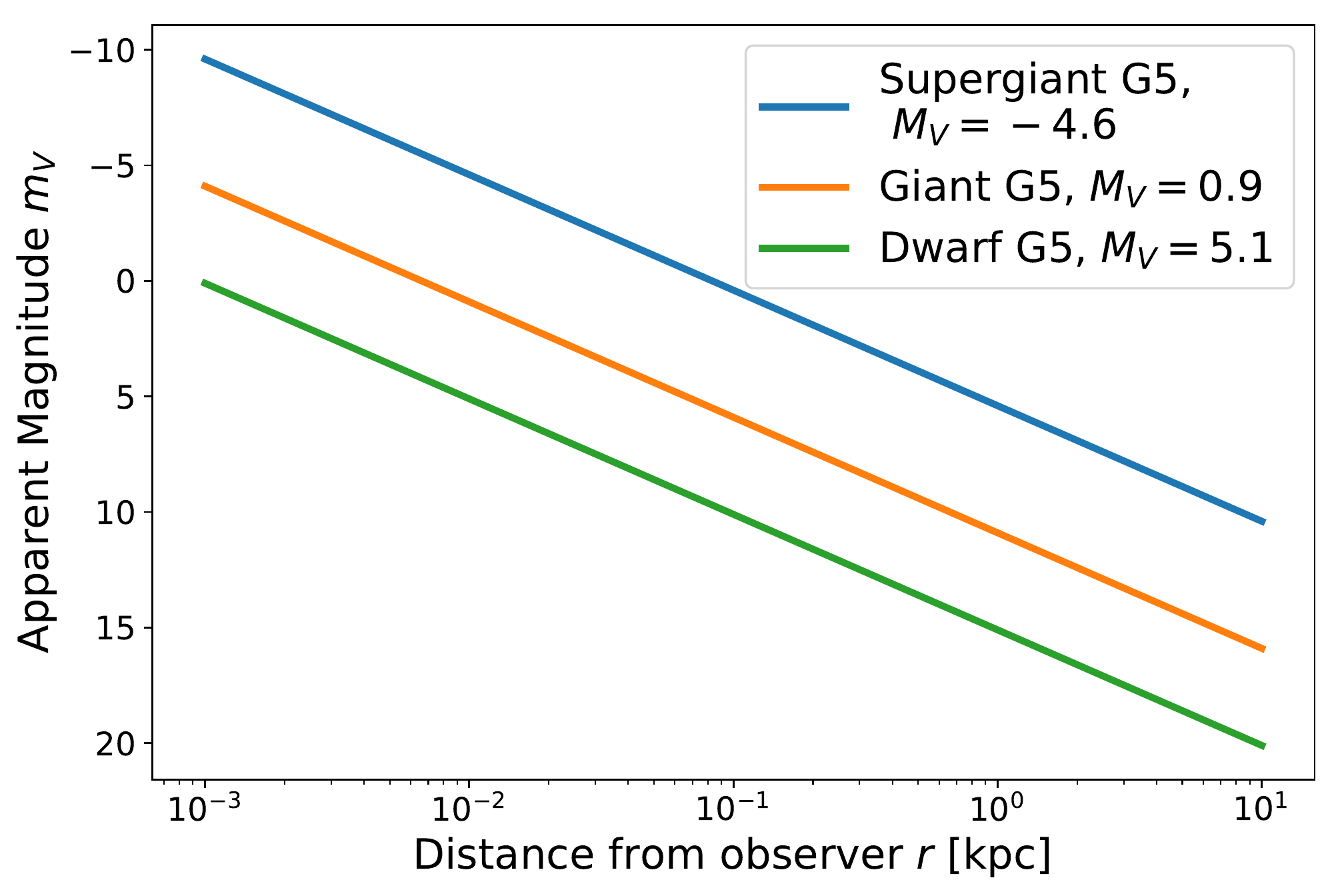}
\caption{Apparent magnitude of G5 dwarfs, giants, and supergiants at various distances from the observer, assuming absolute magnitudes of $M_V = 5.1$, $M_V = 0.9$, and $M_V = -4.6$ respectively \citep{1998gaas.book.....B}, and no reddening or extinction.}
\label{fig:distance_magnitude}
\end{figure}

The cost of using low-jitter dwarf stars is their intrinsic faintness, which limits the distances at which high quality spectra can be obtained. Figure \ref{fig:distance_magnitude} shows the apparent $V$-band magnitude $m_V$ of dwarves, giants, and supergiants of the G5 spectral class over a range of distances $r$, neglecting extinction and reddening, and assuming absolute magnitudes of $M_V = 5.1$, $M_V = 0.9$, and $M_V = -4.6$ respectively \citep{1998gaas.book.....B}. In Figure \ref{fig:mw_accelerations} we saw that to access regions of $\Delta V_{\rm LoS} \sim 1~\cms$ we need to measure stars at distances of around 1 kpc. To derive useful data from a G5 dwarf  at this distance the spectrograph and telescope would need to have a limiting magnitude of  $\gtrsim 15$. For reference, the EXPRES spectrograph that will be installed on the 4.3m Discovery Channel Telescope \citep{2016SPIE.9908E..6TJ, 2017AAS...22912604F} has a limiting magnitude of $m_V \sim 7$, while the ESPRESSO spectrograph, which combines light from the four 8.2m telescopes of the Very Large Telescope (VLT), has a limiting magnitude of $m_V \sim 17$ \citep{2014AN....335....8P, ESPRESSO_website}. Consequently, implementing this technique will require the largest present-day telescopes or next generation instruments. 

On the other hand, giant and supergiant stars are intrinsically brighter and so can be seen at greater distances; ignoring extinction and reddening, a telescope with limiting magnitude  $m_V = 15$ would be able to see a giant star out to $\sim 7$ kpc, reaching into high acceleration regions near the galactic centre. In this region $\Delta V_{\rm LoS}$ values can approach $15~\cms$ over a decade, thus requiring less stringent jitter modelling compared to that needed to extract the $\mathcal{O}(\cms)$ velocity changes expected for stars closer to Sun. Note too that high cadence measurements can reduce the impact of jitter in a long time series, but this will require a trade-off with telescope resources.  

\section{Perspective Acceleration}\label{sec:perspective_acceleration}

\begin{figure}
\includegraphics[width=\columnwidth]{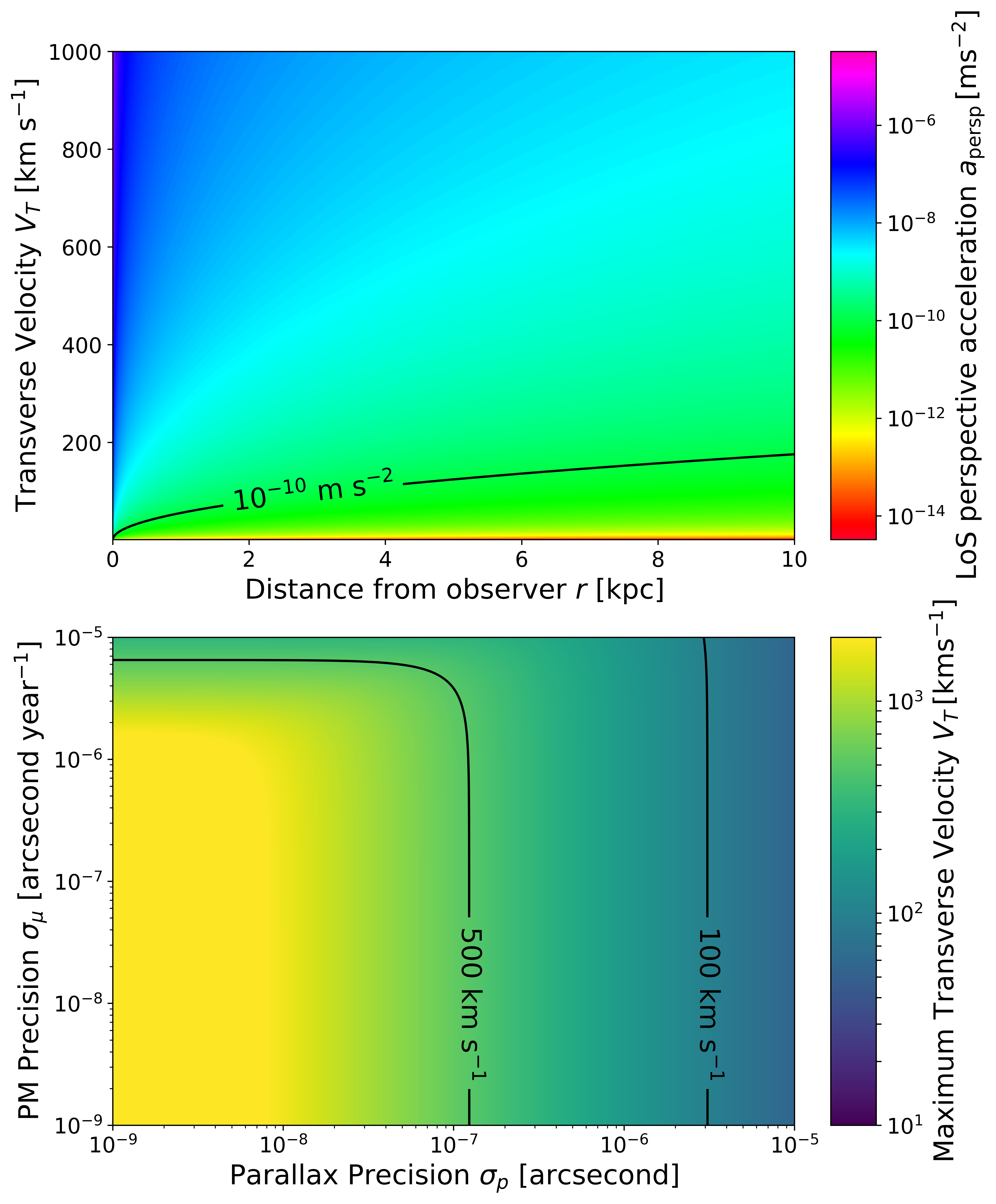}
  \caption{Top: Apparent radial acceleration $a_{\rm pers}$ caused by perspective acceleration for given transverse velocities and distances. Bottom: Maximum $V_T$ compatible with a perspective acceleration absolute error of $\sigma(a_{\rm pers}) = 10^{-12}~\mss$, for varying parallax and proper motion precisions. 
  \label{fig:perspective_acceleration}}
\end{figure}

The apparent radial velocities and proper motions of stars are projections of their velocity vector along the line of sight and the plane of the sky. These projected quantities will change due to relative motion, independently of the acceleration induced by the galactic gravitational field. For a star travelling in a straight line at constant velocity $V$ the instantaneous change in the radial velocity $V_R$ from this effect is \citep{1977VA.....21..289V} 
\begin{align}
a_{\rm persp} = \frac{\rmd V_R}{\rmd t} = \frac{V_T^2}{r} = \frac{V^2}{r_P} \cos^3\theta \, ,
\end{align}
where $V_T$ is the transverse velocity relative to the Sun, $r$ is the distance to the star, $r_P$ is the perihelion distance,\footnote{In this context, ``perihelion'' refers to the distance between the star and the sun at the point of closest approach, where the heliocentric velocity vector  is perpendicular to the line of sight.} and $\theta$ is the angle to the star measured from that point. Clearly the maximum acceleration occurs at perihelion, \ie when $\theta = 0$, $V_T = V$, and $V_R = 0$.  The change in radial velocity from this effect is always positive (when radial velocities away from us are set as positive): stars approaching perihelion have negative $V_R$ that is increasing towards zero, while those moving away from perihelion have positive and increasing $V_R$.  

In the top panel of Figure \ref{fig:perspective_acceleration} we plot the perspective radial acceleration $a_{\rm pers}$ as function of tangential velocity $V_T$ and distance $r$, showing the contour $a_{\rm pers} = 10^{-10}~\mss$, which roughly matches the expected gravitational accelerations shown in the top panel of Figure \ref{fig:mw_accelerations}.    The effect diminishes with distance, but any star within a few kpc of the sun with a relative transverse velocity above $\sim 100~\kms$  has a perspective acceleration comparable to or greater than that expected from the Milky Way potential. 

On the face of it, this  limits us to stars with relatively low  transverse velocities, but sufficiently accurate astrometric measurements will let us correct for this effect. From a rudimentary error analysis, neglecting correlations, the uncertainty on $a_{\rm pers}$ is
\begin{align}
\sigma(a_{\rm persp}) = V_T^2 \sqrt{\frac{ \sigma_p^2}{1 {\rm AU}^2} + \frac{4 \sigma_\mu^2}{V_T^2}}
\label{eq:error_acc_pers}
\end{align} 
where $\sigma_p$ is the uncertainty in parallax (in radians) and $\sigma_\mu$ is the uncertainty in proper motion (in radians per second\footnote{Parallax $\varpi$ is generally measured in arcseconds, and linked to distance via $d~[{\rm pc}] = 1/\varpi$ [arcsecond]. Ostensibly this is dimensionally unbalanced, with distance on one side and inverse angle on the other, but this is due to how a parsec is defined. Thus for ease of unit bookkeeping we revert to the definition of parallax, $\tan{p} = 1~{\rm AU}/r$, where $p$ is the parallax in radians and $r$ is the distance to the star in AU. For the small angles present in astronomy this is approximated as $p \approx 1~{\rm AU}/r$.}). At this level the uncertainty on $a_{\rm pers}$ is independent of distance $r$. Rearranging Equation \ref{eq:error_acc_pers} gives an expression for $V_T$ in terms of $\sigma_p$, $\sigma_\mu$, and the desired level of precision for $\sigma(a_{\rm pers})$
\begin{align}
V_T^2 = \frac{1 {\rm AU}^2}{\sigma_p^2} \left[-2 \sigma_\mu^2 \pm \frac{1}{2}\sqrt{16\sigma_\mu^4 + \frac{\sigma_p^2}{1 {\rm AU}^2} \sigma(a_{\rm persp})^2}\, \right]
\end{align}
As $\sigma(a_{\rm pers}) \propto V_T^2$, we can find an upper limit on the transverse velocities for which we can deduce the perspective acceleration to a desired precision, given $\sigma_p$ and $\sigma_\mu$.  

In the bottom panel of Figure \ref{fig:perspective_acceleration} we plot the maximum transverse velocity compatible with $\sigma(a_{\rm pers}) = 10^{-12}~\mss$ for a range of parallax and proper motion uncertainties. This is a stringent requirement, given that it is 1\% of the typical Milky Way acceleration, but it is not an unattainable one.  The best Gaia precisions are $\sigma_p \sim 10~\muas$ and $\sigma_\mu \sim 10~\muasyr$, yielding a maximum $V_T = 55~\kms$.  Searching the Gaia Data Release 2 catalogue \citep{2016A&A...595A...1G, 2018A&A...616A...1G} shows that there are $\sim 36$ million stars within 1kpc of the Sun and with parallax errors less than 10\%, and $\sim 55\%$ of these have transverse velocities below this threshold. An order of magnitude improvement to $1~\muas$ parallax and $1~\muasyr$ proper motion precision pushes the $V_T$ limit to $175~\kms$ and the percentage of our Gaia DR2 subsample under the $V_T$ threshold to 99\%. Achieving Theia-level precisions of $\mathcal{O}$(100 nas) and $\mathcal{O}(100~\nasyr$) would increase the $V_T$ limit beyond $500~\kms$. 

\section{The Exoplanetary Background}\label{sec:exoplanetary_background}

\subsection{Signals From Exoplanetary Systems}
High precision radial velocity measurements are driven by searches for exoplanetary systems. Ironically, these same exoplanets will be a key source of ``noise'' that must be accounted for and subracted in order to reveal possible gravitational accelerations.  Radial velocity exoplanet searches look for accelerations induced in stars by their orbiting planets, as the star and planet(s) move about their common centre of gravity. For a single planet with period $P$,  the induced velocity oscillation has the {\em semiamplitude}
\begin{align}
K = \left( \frac{P}{2\pi G} \right) ^{-\nicefrac{1}{3}} \frac{M_p \sin i}{M_*^{\nicefrac{2}{3}}} \left(1-e^2\right)^{-\nicefrac{1}{2}} 
\label{eq:semiamplitude}
\end{align}
where $M_*$ and $M_p$ are the masses of the star and planet respectively, $e$ is the eccentricity, $i$ is the inclination of the orbital plane normal vector with respect to the line of sight, with $i=0$ indicating the system is face on \citep{2013pss3.book..489W}. Radial velocity measurements determine the combined term $M_p \sin i$, and thus there is a degeneracy between the mass of the planet and the orbital inclination. 

\begin{figure}
  \includegraphics[width=\linewidth]{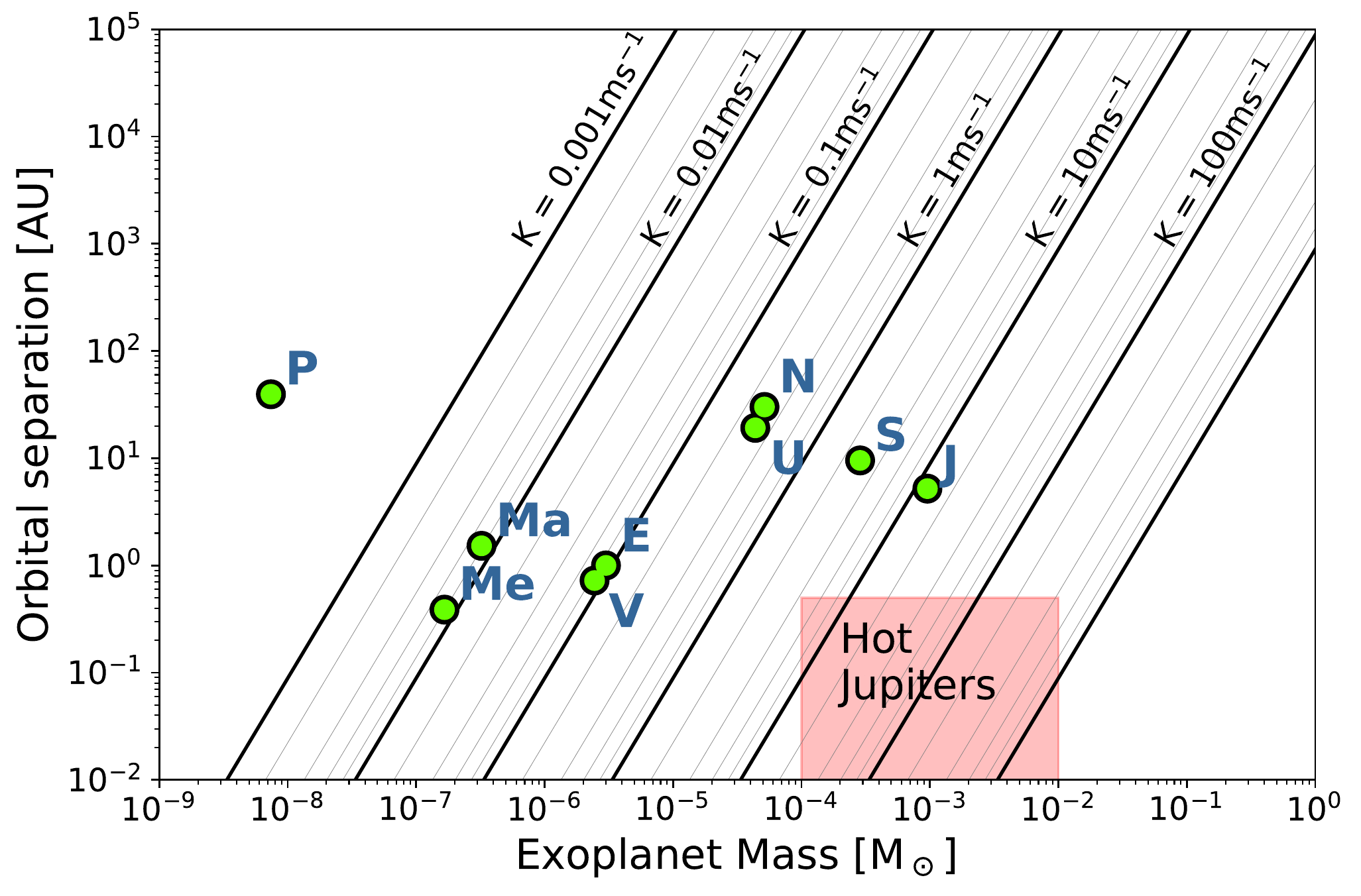}
  \caption{Semiamplitudes $K$ (Equation \ref{eq:semiamplitude}) for a range of exoplanet orbital radii and masses, assuming the host star mass is $M_* = 1~\Msun$, the inclination of the orbital plane is $i = 90^\circ$, and the eccentricity $e = 0$. Also shown are the orbital radii and masses for the eight planets of our solar system and the dwarf planet Pluto, along with a region of radii and masses for the so-called ``Hot Jupiters'' - large gas giants orbiting close to their host stars.}
  \label{fig:K_contours}
\end{figure}

In Figure \ref{fig:K_contours} we show the values of $K$ for a range of planetary orbital radii and masses, assuming a simplified circular ($e=0$) and edge-on ($i=90$) orbit around a host star of mass $M_* = 1~\Msun$, along with the orbital radii and masses of the eight planets and one dwarf planet of our own solar system for comparison. 

 \begin{figure}
  \includegraphics[width=\linewidth]{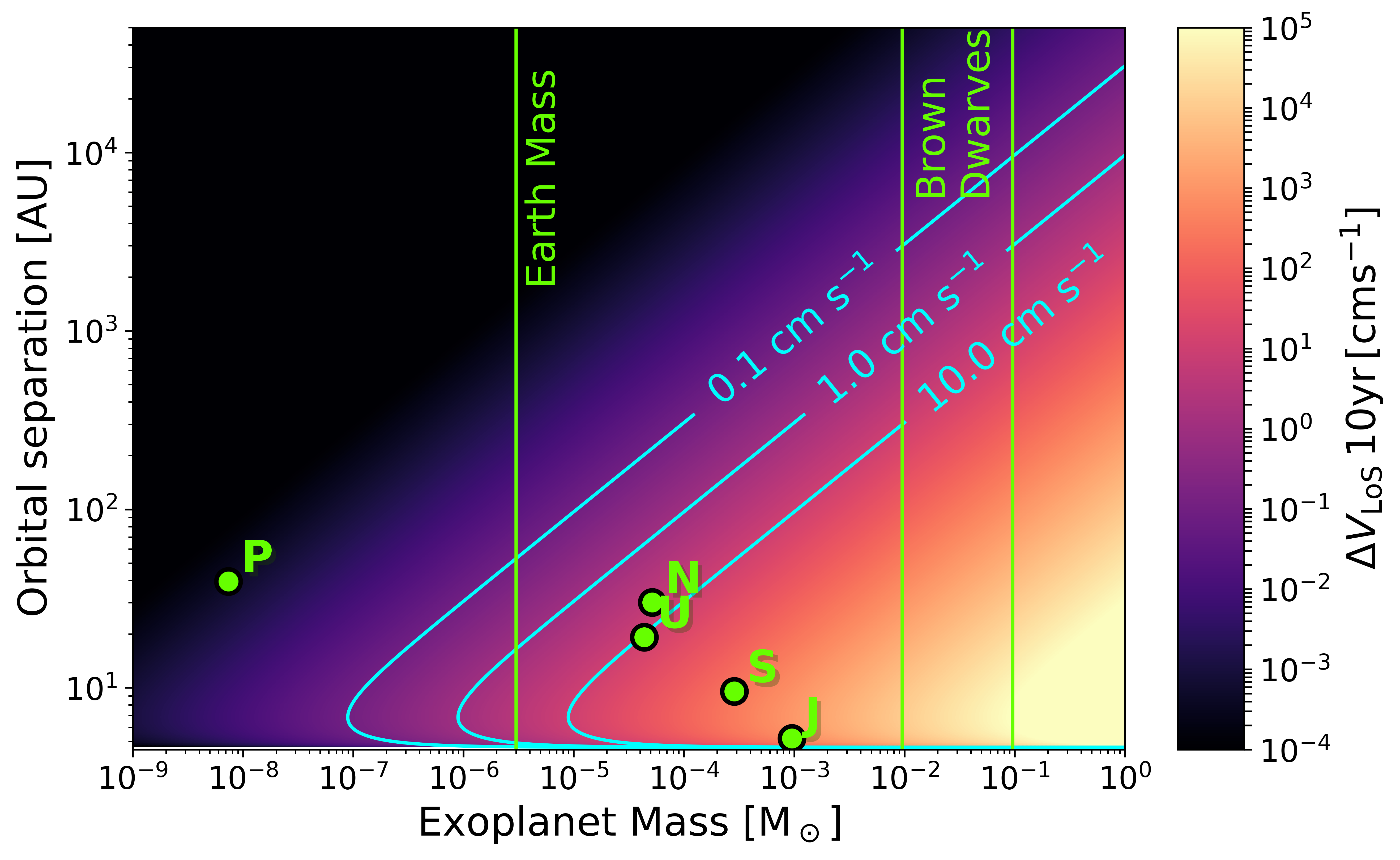}
  \caption{Cumulative change in velocity of a $1~\Msun$ by a single planet of a given mass and orbital radius over a 10 year span. This assumes a circular planetary orbit ($e=0$) that is edge-on to our line of sight ($i=0$). Also shown are the orbital radii and masses for Jupiter, Saturn, Uranus, Neptune, and the dwarf planet Pluto.}
  \label{fig:los_deltaV_exoplanet}
\end{figure}
Assuming this same simplified circular and edge-on orbit, the reflex velocity of a star due to an orbiting planet is then $V_{\rm LoS} = K \cos(\nu(t))$, where $\nu \in [0,2\pi]$ is the angle on the orbital plane, with $\nu = 0$ corresponding to the 3 o'clock position, and increasing counterclockwise. Over a  time interval    $t_1-t_0$ the change in velocity of the star is
\begin{align}
\Delta V_{\rm LoS} = K \left[\cos\left(\nu(t_1)\right) - \cos\left(\nu(t_0)\right)\right] 
\end{align}
To give a qualitative picture of the impact of various planetary configurations, Figure \ref{fig:los_deltaV_exoplanet} shows the cumulative change in velocity over ten years of a $1 \Msun$ star induced by planets of given mass and orbital radius (linked to the period via $P=2\pi \sqrt{R^3/G M_*}$), assuming  the planet is transiting from $\nu(t_0) = \pi/2 - \omega \times 10{\rm yr}/2$ to $\nu(t_1) = \pi/2 + \omega \times 10{\rm yr}/2 $, \eg directly behind the star which maximises the line of sight velocity change\footnote{The initial reflex velocity is zero in this configuration, but the acceleration is maximised.}. For reference we also show the four gas giant planets and the dwarf planet Pluto from our solar system, along with lines indicating the mass of Earth and the mass range for brown dwarfs. We show contours in $\Delta V_{\rm LoS}$  at $0.1~\cms$, $1.0~\cms$, $10.0~\cms$, roughly bracketing the typical $\Delta V_{\rm LoS}$ values generated by the Milky Way potential (see Figure \ref{fig:mw_accelerations}). We exclude planets with periods  of less than 10 years, assuming that planets completing a full orbit in the observation period can be detected, modelled, and marginalised out. 
 
 Gas giants with radii of $\mathcal{O}(1-10 {\rm AU})$, such as Jupiter and Saturn induce large, periodic (or at least time-varying) velocity changes that would be easily detectable. However, large planets with orbital radii in $\mathcal{O}(10-100 {\rm AU})$ range (``Neptunes'') will be more challenging, as they can yield a $\Delta V_{\rm LoS}$ of a similar magnitude to the acceleration to the Milky Way, which will be roughly linear with time on intervals much less than their orbital periods. Likewise Earth mass planets with orbital radii of $\mathcal{O}(10 {\rm AU})$ will contribute similar signals, albeit with a distinct time derivative. 

One might hope to reduce the exoplanetary background by selecting systems with small inclination angle $i$. However the orbital plane would need to be determined with exquisite accuracy: the 10 year $\Delta V_{\rm LoS}$ for a Jupiter mass and period planet at $i=90^{\circ}$ is $\sim 640 \cms$, and only for $i = 0.1^\circ$ does this drop to $\sim 1 \cms$. Furthermore, studies of our own solar system and exoplanetary systems found with \textit{Kepler} suggest mutual inclinations between the planets of $\sim 1^{\circ}-3^{\circ}$ \citep{2014ApJ...790..146F}. Again using the solar system as an example, even if we had the Jupiter-type planet's orbital plane at or below the required $i = 0.1^\circ$ precision, a Saturn mass and period planet with a mutual inclination of $2^\circ$ would produce a $\Delta V_{\rm LoS}$ of $\sim 10 \cms$, overwhelming the Milky Way acceleration.

\subsection{Reducing the Exoplanetary Background}\label{sec:Filtering_Exoplanetary_Background}

We  focus on broad strategies that have the potential to reduce or exclude the exoplanetary signal, leaving the detailed data analysis challenge associated with extracting the galactic acceleration signal from radial velocity data to future work\footnote{The contemporaneous paper by  \citep{HarvardMichiganPaper} looks  explicitly at this.}.  We can see three specific approaches: identifying systems for which this background is likely to be intrinsically small; combining radial velocity information with other data on planetary systems to improve the fitting process; and combining spatially adjacent stars to separate the correlated acceleration arising from the galactic potential from the largely random residual accelerations caused by exoplanets. 

\subsubsection*{Single System Measurements}\label{sec:single_system_measurements}
If we can identify classes of stars which are substantially less likely to have exoplanets (or at least those in the most problematic categories identified above) the $\Delta V_{\rm LoS}$ due to the Milky Way potential can be more easily and reliably extracted. Given the timeframe in which this experiment is likely to be conducted, it is possible that signatures (based on detailed elemental abundances, spectral type and dynamical properties) of stars with complex planetary systems will be more clearly quantified than is the case at present. 

Stars moving with very high speeds relative to the Milky Way restframe may be a fruitful category to consider, as they may  be born without planets, or be stripped of their planets by the mechanism responsible for  accelerating them.  These stars come in several subcategories, as delineated by \citet{2017MNRAS.469.2151B}. \textit{Hypervelocity stars} (HVSs) are stars with sufficient velocity to be unbound from the Milky Way, regardless of production mechanism; this is roughly $\sim 500~\kms$ in the solar neighbourhood, rising to $\sim 600~\kms$ in the galactic centre \citep{2017MNRAS.468.2359W,2018A&A...616L...9M}. \textit{Hills stars} are accelerated via a three-body interaction of a tight binary with the central black hole of either the Milky Way or the LMC, a process called the \textit{Hills} mechanism \citep{1988Natur.331..687H}, while \textit{runaway stars} are accelerated by the supernova explosion of their binary companion \citep{1961BAN....15..265B}. 

Runaway stars \citep{1961BAN....15..265B}  are posited to form from binary system where the larger star sheds much of its mass in a supernova explosion, releasing the less massive star at its (potentially high) orbital velocity.  Velocities can be up to $\sim 200~\kms$ in the galactic rest frame and point back to where the supernova explosion took place. This ejection mechanism may not strip the star of its exoplanets, but stars that form in relatively tight binaries are possibly less likely to possess long-period planets to begin with. 

Stars can also be accelerated to high velocities through dynamical encounters in dense clusters - these fall outside the categorisation of \citet{2017MNRAS.469.2151B} but can be referred to as dynamical ejection scenario (DES) runaways \citep{2001A&A...365...49H}. The bulk of DES runaways are actually \textit{walkaway} stars, with velocities $< 30~\kms$ \citep{2011MNRAS.414.3501E, 2018arXiv180409164R}. Whether the DES process can remove any planets requires further investigation, but would likely depend on the proximity of the encounters.   

The Hills mechanism accelerates stars via the interaction of a tight binary with a massive black hole, with one star being captured by the black hole and the other ejected at velocities of $\mathcal{O}(1000~\kms)$. For Hills stars in the Milky Way the black hole responsible would be Sagittarius A$^*$ \citep{1988Natur.331..687H}, or possibly a massive black hole in the Large Magellanic Cloud \citep{2016ApJ...825L...6B}.  

The fate of planetary systems around Hills stars was analysed by \citet{2012MNRAS.423..948G}, who performed simulations where both members of the binary  have exoplanets,  finding that hypervelocity Hills stars can retain  planets. However, these were typically very tightly bound to the parent star with orbital radii in the range of 0.02-0.05 AU. At smaller radii planets will interact with the star itself; at larger radii  dynamical instabilities eject  planets or cause them collide with one of the stars before the final interaction with the black hole. Thus, the surviving planets  have very short periods making their radial velocity signal easier to extract.   

Initially theorised in 1988, the first hypervelocity Hills star was discovered in 2005 \citep{2005ApJ...622L..33B}. This star, HVS1, is a B-type star with an effective temperature of $T_{\rm eff} \sim 10500$ K, making it far too hot for  precise radial velocity measurements (see Section \ref{sec:spectroscopic_performance}).  Other candidates are likewise hot, massive, short lived stellar types, as they are easily spotted in a background  of old halo stars  \citep{2006ApJ...640L..35B, 2007ApJ...671.1708B}. However nothing appears to preclude older hypervelocity Hills stars - they could in fact be more numerous but simply harder to find in the halo  \citep{2007ApJ...664..343K, 2010ApJ...723..812K}. 

Several features allow Hills stars to be selected from stellar catalogs. Their velocity vectors should (roughly) point away from their origin, the Galactic Centre. Secondly, since Hills stars come from the Galactic Centre they should have the high metallicity characteristic of their origin. For instance \citet{2016ApJ...832...10Z} found 29 F- and G-type low mass stars with orbits and high metallicities consistent with Galactic Centre origins, though none had velocities above the escape speed and so did not qualify as HVSs. However, such stars may still be suitable for our purposes if they are above the threshold required to remove problematic long period exoplanets.
  
As discussed  in Section \ref{sec:perspective_acceleration} HVSs necessarily have large perspective accelerations; using these stars as probes of the galactic magnetic field will need either  increased astrometric precision to  calculate their proper motion and parallax and thus their radial perspective acceleration, or a further cut to remove stars that were not moving in directions closely aligned to the line of sight, reducing their transverse velocity. 

Finally, a separate approach would be to focus on relatively tight binaries -- in such systems planetary systems are more likely to be disrupted, reducing the likelihood of finding the sort of ``Neptunes'' that would potentially contribute the most challenging radial velocity changes.  However, these systems would likely be spectroscopic binaries, increasing the difficulty of extracting the underlying radial velocity. 

\subsubsection*{Combining  Exoplanet Detection Techniques}
A different strategy would be to target stars for which complementary exoplanet detection methods had (at least partially) mapped their planetary systems, providing additional constraints on the fit to the radial velocity data and so facilitating the extraction of the Milky Way acceleration signal. Critically, this  includes null results that would exclude possible planetary candidates.  In particular, it would valuable to rule out the presence of long period/large radius Neptunes ($\gtrsim 10$ years, $\gtrsim 5$ AU), and to break the $M_p \sin i$ degeneracy between exoplanetary masses and radial accelerations.  

Direct imaging searches are ideally suited for detecting large planets at large radii, and thus long periods, for the obvious reason that this maximises the contrast and angular separation on the sky between star and planet \citep{2013pss3.book..489W}. Direct imaging at multiple epochs can also break the $M_p \sin i$ degeneracy \citep{2017AJ....153..135M,2016ApJ...832...33H}. Several direct imaging instruments are currently operational, such as the Gemini Planet Imager \citep{2006SPIE.6272E..0LM}, VLT-SPHERE \citep{2008SPIE.7014E..18B}, and Subaru Coronagraphic Extreme Adaptive Optics (SCExAO) \citep{2015PASP..127..890J} which is also serving as a technology testbed for exoplanet imaging with the Extremely Large Telescope (ELT) \citep{2013aoel.confE..94J, 2018arXiv181002031C}. There are also several planned and proposed space missions to conduct direct imaging, such as WFIRST \citep{WFIRST_website,2016JATIS...2a1001N}, and the New Worlds Observer \citep{2010SPIE.7731E..2EL}.  However, to be useful these imaging studies would need to be sensitive to planets at kiloparsec distances. 

Gravitational microlensing is another method well suited to detecting exoplanets at larger radii \citep[\eg][]{2018Geosc...8..365T} and hence is highly complementary. As a foreground lens object transits the line of sight between a background source and an observer the magnification of the source will smoothly increase, peak, and then decrease symmetrically in time. If the lens object hosts an exoplanet it too can focus the light of the background star, adding a secondary off-centre peak in the magnification curve \citep{1991ApJ...374L..37M, 1992ApJ...396..104G}. Microlensing  requires an element of luck as the lens star and planet have to cross sufficiently close to the line of sight to the background source, but known lenses could provide an interesting set of target stars. 

Astrometric detection could also break the $M_p \sin i$ degeneracy. This method relies on the same reflex motion of the star as the radial velocity technique, but is sensitive to the orthogonal velocity component, which induces 2D proper motion variations on the plane of the sky. Combining with radial velocities to yield a full 3D determination of the velocity variation allows for a determination of the inclination of the orbital plane, breaking the $M_p \sin i$ degeneracy \citep{2009ApJS..182..205W, 2013pss3.book..489W}. For a fixed perpendicular velocity the measured proper motion decreases with distance, meaning that this technique is most effective with nearby stars. For example, a perpendicular velocity of $1\ms$ would translate to a proper motion of  $\sim 150 \muasyr$ at 1.3 pc (\ie the distance to Alpha Centauri), but $\sim 200~\nasyr$ at 1 kpc.  Changes in proper motions between the Hipparcos catalogue and Gaia DR2 measurement have been used to measure $\mathcal{O}({\rm mas})$ accelerations of a handful of G- and K-dwarf stars and so detect the presence of brown, white, and M-dwarf companions \citep{2018arXiv181107283B, 2018arXiv181107285B}.

These complementary techniques could be linked by constructing models for each exoplanetary system, potentially with additional priors from planetary formation and dynamics models, forward modelling into the data space of each experiment, and simultaneously fitting to all available data. Such a fit would yield a measure of the maximum radial acceleration attributable to unseen exoplanets, and the residual acceleration would be attributable to the Milky Way potential.

\subsubsection*{Multi-Star Techniques}
The most obvious approach is to combine observations of multiple stars, under the assumption that the alignments of their exoplanetary orbital planes and phases, and hence the resultant acceleration vectors, are random and uncorrelated, whereas the acceleration from the Milky Way potential is coherent over large distances. A rudimentary implementation would be to simply stack radial velocity measurements for adjacent stars, in combination with models for the likely exoplanetary background.  

A more sophisticated route would be to add a common acceleration term to the multi-experiment models described above, and conduct a simultaneous fit across multiple stellar systems. A correlated residual acceleration across many stars would be more likely to come from the Milky Way acceleration field than a conspiracy of perfectly aligned exoplanets sitting below the detection threshold of the complementary experiments. We save a full treatment of this topic for future work, but note that the next decade(s) will see an explosion of information about the global properties of exoplanetary systems.

\section{Galactic Gravitational Field} \label{sec:disc_morphology}
Having considered the  challenges one would encounter in performing these measurements, we  finally turn to examining specific  investigations of the galactic gravitational field that would be facilitated by high precision radial velocity measurements. Beyond the possibility of gaining a new set of constrains on models of the galactic gravitational field and the underlying mass distribution in the Milky Way, we discuss two  opportunities;  searches for localised over-densities in the (assumed) dark matter distribution and tests of modified gravitational dynamics.

\subsection{Subhalos and Overdensities} \label{sec:overdensities} 
Many scenarios admit the possibility of large inhomogeneities in the dark matter distribution. These include the remnants of halos associated with mergers that have yet been fully disrupted and assimilated, or exotica such as ultracompact minihalos \citep{2009ApJ...707..979R,2012PhRvD..85l5027B} and the local inhomogeneities that might arise in models such ultralight (or fuzzy) dark matter \citep{2017PhRvD..95d3541H}. A large, localised acceleration that was not matched by a similar anomaly is the distribution of baryonic matter would be clear evidence for dark matter -- but the existence of such inhomogeneities  depends on both the detailed properties of the dark matter  and the dynamical history of the Sun's neighbourhood. 

To investigate this, consider an idealised spherical overdensity. At a distance $r$ from (and external to) a spherical object of mass $M$  the acceleration it induces is
\begin{align} \label{eq:accel}
a =  1.4 \times 10^{-13} \frac{M}{\Msun}  \left( \frac{1 \textrm{pc}}{r} \right)^2 ~\mss \, .
\end{align}
The obvious observational strategies are  simple; stars in the vicinity of an isolated overdensity will have an additional, anomalous acceleration. For stars on a line of sight through the centre of the overdensity, the total range of the modulation is twice the value by given Equation ~\ref{eq:accel}; the acceleration has the opposite sign on opposite sides of the overdensity. 

Standard cold dark matter (CDM) cosmology suggests that the  Milky Way halo includes many subhaloes. Typical masses and sizes for these objects are in the range of $10^7 \Msun$ and a radius of 250 pc to $10^8 \Msun$ with a radius of 625 pc \citep{2015MNRAS.454.3542E}. These would produce a 10 year velocity change of 1.4  and 2.3 $\cms$ respectively across the object, which is likely to be experimentally accessible. The variation as a function of angular direction would be much larger than that of the background field if the object is at a distance from the sun at least several times larger than its radius. Much denser objects are also frequently discussed --  to give one example,  \citet{2018arXiv181103631B} analysed the dynamics of stellar streams,  positing the existence of objects with radii of just 20 pc and masses between $10^{5.5}$ and $10^{8}~\Msun$, with corresponding 10 year velocity changes  from $7$ to $\sim2000~\cms$. The latter signal is well with the reach of present-day instruments, albeit for stars in a relatively small spatial volume in the vicinity of the object. 

This analysis is at the proof-of-concept level, but it is sufficient to demonstrate that plausible compact subhalo objects would be detectable via their contributions to stellar accelerations. 

\subsection{MOND v Newton} \label{sec:mond_newton}
Theories of Modified Newtonian Dynamics (MOND) \citep{1983ApJ...270..365M,2010MNRAS.403..886M} postulate that galaxies are composed of baryonic matter and that their global dynamics are explained by modifications to Newton's laws that become manifest at large distances and/or very small accelerations. While  $\Lambda$CDM \citep[\eg][]{2018arXiv180706205P} is well-established as the standard model for cosmology, modified gravity theories can remain  viable in the absence of a definitive identification of the physical basis for dark matter. 

Because MONDian theories are designed to mimic the rotational dynamics of spiral galaxies such as the Milky Way,  their predictions will typically overlap with Newtonian gravity in presence of dark matter for motion in the galactic plane. However, the baryonic matter has a very different distribution from the putative dark  component -- the former is roughly axisymmetric, whereas the dark halo is (approximately) spherical. 

Given these differing source geometries it would be surprising if dynamical predictions of the two theories overlapped globally,  and in fact they do not. The effective gravitational potential in a MONDian Milky Way is discussed by  \citet{2008MNRAS.386.2199W} and \citet{2009A&A...500..801B}.  In these scenarios, the Milky Way acquires an apparent ``phantom disc'' if the observed dynamics in the $z$ direction are solved assuming that the underlying dynamics are Newtonian, and the direction of the acceleration vector likewise differs between these scenarios. This  leads to differing predictions for the  vertical accelerations as a function of height; allowing observations to break the degeneracy between the two scenarios. Jeans equation based techniques have already been used to fit acceleration maps to position and velocity data, which were then used to test MONDian against Newtonian dynamics \citep{2014ApJ...794..151L}.
 
One potential observational strategy would be to first select stars that were co-rotating with the Sun via a proper motion cut to reduce perspective effects, and then measure the accelerations vertically above and below the plane ($z$-direction). This would mean $a_z \approx a_{\rm LoS}$, with a reduced degeneracy in proper motion accelerations, allowing for a simple test of the vertical acceleration prediction derived from the MONDian ``phantom disc'' against the spherical CDM halo.

With full radial velocity and proper motion acceleration data the potential of the Milky Way could be derived using the Poisson equation alone in Newtonian gravity. However models could be fit to even sparse radial acceleration information while marginalizing over the unknown proper motion accelerations, and with the use of highly constraining acceleration data these models could be very general and have few assumptions. 

\section{Conclusions and Discussion} \label{sec:conclusions}
Our investigation of a typical model of the galactic potential shows that gravitational accelerations will lead to changes in the line of sight velocities of stars on the order of centimetres per second over a decade.  Given that the advent of $\cms$ spectrographic precision is  approaching, it may soon be possible to directly measure the acceleration field of the Milky Way. 

That said, many challenges  need to be overcome to realise such a measurement.  Perhaps the biggest unknown is our ability to exclude jitter -- induced by motions in stellar atmospheres -- which can generate spurious radial velocity signals of $\mathcal{O}(100\cms)$ for dwarf stars, and $\mathcal{O}(10~\ms)$ for giants. The processes responsible for jitter modify the detailed spectra, allowing it to be distinguished from the pure Doppler shifts associated with the mean stellar velocity \citep[\eg][]{2017ApJ...846...59D} and very high dispersion and resolution spectrographs coupled to large telescopes may permit further improvements. The need to exclude jitter may  guide the selection of target stars. Here we face a trade-off: dwarf stars have lower levels of jitter but are inherently fainter, and observations of them may be limited to regions of the galaxy relatively close to the Sun where accelerations are expected to be lower. Conversely, giant stars have higher jitter but are brighter and thus observable further into the high acceleration regions close to the galactic centre. 

As explained in Section \ref{sec:perspective_acceleration}, precise astrometric observations will be crucial to measurements of the galactic acceleration field. This is due to perspective acceleration, where purely geometric effects associated with proper motion also change stellar radial velocities. However, while this apparent acceleration is larger than the gravitational acceleration, it can be calculated and subtracted if the parallax and proper motion are known to sufficient precision.  Excluding perspective acceleration to the point that it is less than 1\% of the characteristic Milky Way acceleration with Gaia-level astrometric  precision restricts us to stars with relative transverse velocity of less than $55~\kms$, but roughly half of all local stars survive this cut. An improvement of an order of magnitude in astrometric precision would make most local stars usable as probes of the galactic acceleration field.  If astrometric precisions reach the nano-arcsecond level, perspective accelerations would be trivially removable. Further, as we saw in Section~\ref{sec:Milky Way_acceleration_field}  such measurements would  yield the transverse component of the gravitational acceleration induced by the Milky Way, allowing for direct mapping of the Milky Way's gravitational field and mass distribution when combined with radial accelerations.

Exoplanet searches have motivated the improvement of spectrographic precisions, and ironically exoplanets will be a key source of noise when measuring the Milky Way's acceleration field. Accelerations from short period planets will be modulated over time, making it possible to extract them, However,  large planets with long orbital periods  (``Neptunes'') would yield  accelerations of similar magnitude to those of the Milky Way, with a roughly linear change in velocity over a period of a decade.  We discussed several possible strategies for dealing with this issue.  These include working with stars that have been accelerated to high velocities, as the process that accelerated them is likely to have removed any long period planets, or prevented  their formation in the first place. Similarly, binary systems may be less likely to host long period planets, although they would present a greater spectroscopic challenge. Next, complementary detection techniques such as direct imaging or microlensing may reveal additional information which can better constrain the possible exoplanetary contribution to any residual change in the radial velocity. Moreover, a better general understanding of the possible ``architectures'' of extrasolar planetary systems may further facilitate this analysis. Finally, we can combine measurements of multiple systems, on the assumption that exoplanet orbital planes and phases will be randomly oriented, while the Milky Way's acceleration field will vary smoothly throughout space. 

While substantial advances in radial velocity and astrometric precision, jitter modelling, and exoplanetary background reduction are needed to detect accelerations arising from the Milky Way's gravitational potential, the scientific reward of these measurement would be significant. The potential and mass distribution of the Milky Way could be studied without relying on the assumptions required to extract these quantities from positions and velocities with methods such as distribution function or Jeans modelling. 

There are many possible dark matter scenarios that are effectively impossible to verify directly. In fact, it could plausibly be argued that such scenarios are actually the norm, given that vast range of models that can be imagined and the absence of any truly meaningful guidance from fundamental theory as to the nature of dark matter.  Consequently, despite the large, ongoing experimental effort for specific models, ``blind tests'' could be critical to definitively testing the dark matter hypothesis. These would need to be searches for fingerprints of dark matter that are independent of its specific properties, and focus instead on the dynamics of stars and galaxies and gravitational physics at kiloparsec length scales.  

Acceleration measurements could definitively establish the presence of the extended, roughly spherical dark matter halo predicted by the $\Lambda$CMD model. Furthermore, the presence of dark matter subhaloes is a key signature of the hierarchical structure formation that is natural in $\Lambda$CMD models. These and other types of dark matter overdensities would generate additional accelerations above that of the total Milky Way potential, and would be visible in the radial acceleration distribution as crescents perpendicular to the line of sight. 

A direct measurement of acceleration at galactic scales would also be a powerful test of modified gravity theories. Typically these theories find their strongest support in galactic dynamics \citep{2012LRR....15...10F}, rather than phenomena with larger characteristic comoving scales, such as the cosmic microwave background and large scale structure \citep[\eg][]{2004ApJ...604..596C, 2005PhRvD..72j1301S, 2006ApJ...648L.109C, 2011IJMPD..20.2749D, 2015PhRvD..92h3505X}.  However, galaxies are strongly driven by nonlinear dynamics and baryonic feedback, so determining the detailed predictions of competing theories in this regime is inherently difficult. However, direct measurements of accelerations inside our galaxy  have a unique capacity to test modified gravity theories in the settings where their proponents find them the most persuasive. 

So what would an acceleration measurement campaign look like? The spectrograph and telescope system would need to have a limiting magnitude of $m_V \gtrsim 15$ in order to observe dwarf stars in regions with $\Delta V_{\rm LoS} \gtrsim 1 \cms$. The ESPRESSO spectrograph at the VLT has a limiting magnitude of $m_V = 17$, and  any effort to measure the Milky Way acceleration would require a telescope of a similar class or larger.  

Given that multiple observations of multiple stars will be needed over a decade, a well-designed campaign would be critical to success.  In the shorter term we could imagine carefully investigating improved techniques for understanding jitter with an instrument like ESPRESSO, as a warmup for a bigger survey.  The issues associated with the extraction of planetary motion are discussed in \citet{HarvardMichiganPaper}. Carefully designed observing strategies that take full advantage of complementary information could make this sort of survey vastly more tractable than a  ``brute force'' analysis, given that accuracy gains will ultimately run as the square root of the total number of stars.  Similarly, there will be trade-offs between making  high cadence observations of a small number of stars, verses lower cadence observations of many, and survey design will depend strong on specific science goals. 

The proposal discussed here is  an ambitious one, and its  full realisation would take  decades rather than years. However, our analysis shows that obtaining direct measurements of  accelerations and  net gravitational forces at galactic distances is achievable with a reasonable extrapolation of present day technology, and would create new and exciting opportunities for  galactic astronomy, astrophysics, and fundamental physics. 

\begin{acknowledgements}
 
H.S. acknowledges support from the MINECO (Spanish Ministry of Economy) - FEDER through grant MDM-2014-0369 of ICCUB (Unidad de Excelencia `Maria de Maeztu').  R.E. acknowledges  support from the Marsden Fund of the Royal Society of New Zealand. HS would like to thank \mbox{Monica} Silverwood and Rob Dixon for their hospitality during the early development of this project. We thank Tristan Cantat-Gaudin, JJ Eldridge, Claus Fabricius, Anna Ferr\'{e} Mateu, Francesca Figueras, Debra Fischer, Mark Gieles, Jordi Miralda-Escude, Roger Mor, Pau Ramos, Nicholas Rattenbury, Justin Read, Merc\'{e} Romero-G\'{o}mez, Daniel del Ser Badia, and Willem van Straten for useful conversations. This work has made use of data from the European Space Agency (ESA) mission {\it Gaia} (\url{https://www.cosmos.esa.int/gaia}), processed by the {\it Gaia} Data Processing and Analysis Consortium (DPAC, \url{https://www.cosmos.esa.int/web/gaia/dpac/consortium}). Funding for the DPAC has been provided by national institutions, in particular the institutions participating in the {\it Gaia} Multilateral Agreement. 
\end{acknowledgements}

\begin{appendix}
\section{Spectrographic Precisions Through History}\label{app:spectrograph_references}
The following spectrographs and measurements were included in Figure \ref{fig:spectrograph_precision}:
\textit{Campbell \& Walker} \citep{1988ApJ...331..902C};  
\textit{Mt Wilson/Hamilton 1989} \citep{1989ApJ...344..441M};  
\textit{CfA/CORAVEL} \citep{1989Natur.339...38L};  
\textit{Hamilton 1992} \citep{1992PASP..104..270M};  
\textit{McDonald Observatory Planetary Search} \citep{1993ApJ...413..339H};  
\textit{ELODIE} \citep{1995Natur.378..355M, 1996A&AS..119..373B};
\textit{Hamilton 1996} \citep{1996PASP..108..500B};
\textit{HIRES} \citep{1994SPIE.2198..362V};
\textit{UCLES-AAPS} \citep{2001ApJ...555..410B};
\textit{Tull} \citep{1995PASP..107..251T};
\textit{HRS} \citep{1998SPIE.3355..387T};
\textit{HERCULES} \citep{2002ExA....13...59H, 2015IJAsB..14..305E};  
\textit{HARPS} \citep{2003Msngr.114...20M, 2002Msngr.110....9P}; 
\textit{PFS} \citep{2006SPIE.6269E..31C};
\textit{SOPHIE+} \citep{2008SPIE.7014E..0JP,2013A&A...549A..49B}; 
\textit{CHIRON} \citep{2013PASP..125.1336T};
\textit{HARPS-N} \citep{2014SPIE.9147E..8CC, 2013A&A...554A..28C};
\textit{PARAS} \citep{2010SPIE.7735E..4NC, 2014PASP..126..133C};
\textit{Levy CPS} \citep{2010SPIE.7735E..4KR};
\textit{Levy LCPS} \citep{2010SPIE.7735E..4KR, 2014PASP..126..359V};
\textit{SONG} \citep{2011JPhCS.271a2083G};
\textit{ESPRESSO} \citep{2010SPIE.7735E..0FP, 2014AN....335....8P, ESPRESSO_website};
\textit{EXPRES} \citep{2017AAS...22912604F};
\textit{G-CLEF} \citep{2012SPIE.8446E..1HS, 2018SPIE10702E..1RS};
\textit{CODEX}, \citep{2010SPIE.7735E..2FP, 2010Msngr.140...20P}.
\end{appendix}

\bibliographystyle{pasa-mnras}
\bibliography{acceleration_bibliography}

\end{document}

%% file: jdefs.tex
\def\aj{AJ}%
\def\araa{ARA\&A}%
\def\apj{ApJ}%
\def\apjl{ApJ}%
\def\apjs{ApJS}%
\def\ao{Appl.\ Opt.}%
\def\aap{A\&A}%
\def\mnras{MNRAS}%
\def\prd{Phys.\ Rev.\ D}%
\def\pasp{PASP}%
\def\nat{Nature}%